\documentclass[runningheads]{llncs}
\usepackage[numbers,sort&compress]{natbib} 
\usepackage{graphicx}
\usepackage{algorithm}
\usepackage{booktabs,caption,threeparttable}
\usepackage{algpseudocode}
\usepackage{multirow}
\usepackage{xcolor}
\usepackage{url}
\usepackage{tikz}
\usepackage{amsmath}
\usepackage{pdflscape}

\begin{document}
\title{Privacy-Preserving Billing for Local Energy Markets}
%
%
\author{Eman Alqahtani\inst{1,2} \and
Mustafa A. Mustafa\inst{1,3}
}
\authorrunning{E. Alqahtani and M. Mustafa}
%
\institute{The University of Manchester, Manchester, UK 
\\ \and
FCIT, KAU, Jeddah, 1589, Saudi Arabia \\ \and
COSIC, KU Leuven, Leuven, 3001, Belgium \\
\email{eman.alqahtani@postgrad.manchester.ac.uk}
}
\maketitle              
\begin{abstract}
We propose a privacy-preserving billing protocol for local energy markets (PBP-LEM) that takes into account market participants' energy volume deviations from their bids. PBP-LEM enables a group of market entities to jointly compute participants’ bills in a decentralized and privacy-preserving manner without sacrificing correctness. It also mitigates risks on individuals’ privacy arising from any potential internal collusion. We first propose an efficient and privacy-preserving individual billing scheme, achieving information-theoretic security, which serves as a building block. PBP-LEM utilizes this scheme, along with other techniques such as multiparty computation, inner product functional encryption and Pedersen commitments to ensure data confidentiality and accuracy. Additionally, we present three approaches, resulting in different levels of privacy protection and performance. We prove that the protocol meets its security and privacy requirements and is feasible for deployment in real LEMs: bills can be computed in less than five minutes for 4,000 users using the most computationally intensive approach, and in just 0.18 seconds using the least intensive one.
\end{abstract}

\section{Introduction}
In the quest for a sustainable future, local energy markets (LEMs) are emerging as transformative factors, with several projects already implemented~\cite{ZHANGCH2017}. These markets promote the adoption of renewable energy sources and support local generation/consumption, reshaping the energy system to become more resilient and community-driven, which offers a promising plan to achieve net-zero targets. Their goal is to allow individuals with renewables to directly trade their surplus energy with others in open markets to optimise social welfare while maintaining local balance~\cite{Shafie-khah}. This contrasts with the traditional seemingly unfair practice of limiting individuals to selling excess energy to their contracted suppliers at a fixed feed-in-tariff (FiT), that is much lower than retail market prices (RP)~\cite{FeedINTariffs}.

LEMs typically require their participants to submit bids before the actual trading periods (ranging from 30
minutes to a day in advance)~\cite{Timothy}. Therefore, market participants need to forecast the necessary bid volumes, indicating the amount of energy they intend to trade. This prediction relies on historical data and estimated consumption, making it inherently prone to errors and challenging to achieve absolute accuracy. Whether due to intentional actions or inaccuracies in predictions, participants may commit to trading specific energy volumes but subsequently fail to fulfill these commitments. 

As this failure can disrupt grid stability and increase balancing costs, market participants should be accounted for their deviations from their committed bid volumes during the billing process~\cite{Madhusudan}. This consideration would incentivise the participants to reduce their deviations and, subsequently, the negative effect they cause on the grid. Furthermore, a market participant's deviation may cause a different level of effect on the grid depending on what part of the grid the participant resides in. As a result, some might bear higher costs for their deviations compared to others, not only due to their deviation amounts but also because of their specific locations on the grid. For example, the total deviation of households (i.e., market participants) connected to a particular distribution line may result in a total deviation of zero. That means the positive deviations of the households on that line compensate for the negative deviations. As such, it is rational not to account those individuals for their deviations regardless of the total deviation of the entire market area.

The computation and settlement of bills and rewards for participants in LEMs, while considering their energy deviations from their bids, requires critical private information, such as individual bid volumes and meter readings for each trading period. Such information are closely associated with individuals' consumption patterns, which have been shown to pose potential risks to their privacy~\cite{Quinn}. For example, personal details such as presence at home, sleep patterns or even the detection of particular appliance activation could be inferred from the amount of energy traded at a particular period~\cite{Quinn,Jawurek}.

A number of previous studies have already addressed the privacy concerns of market participants during billing and settlement, employing techniques such as anonymisation~\cite{ZhangX,Dorri,Dorri2,Radi,Baza,Aitzhan,WangY,Dimitriou,LiS,Eisele,Aron2}, perturbation~\cite{Gai,Xiaoyan,Hassan}, and homomorphic encryption~\cite{Kamil,Andrei}. However, the majority have primarily suggested payment approaches based only on bid commitments, assuming perfect fulfillment of the bid volumes. A very limited number of the privacy-preserving billing studies have considered establishing a billing mechanism based on the actual amount of produced/consumed energy by the market's participants~\cite{Xiaoyan,Gai,Gaybullaev,Son} or to additionally account for their deviations~\cite{Kamil,Andrei,Eman}. However, the billing process is coordinated or executed by a single party~\cite{Xiaoyan,Gai,Andrei,Gaybullaev,Son}, assuming trustworthiness and honesty from that party in performing the billing process, disclosing some or all of the individuals private data to that trusted party and potentially imposing a high computational cost on a single party. All of the above raise questions about the applicability or scalability of their solutions in realistic scenarios. In contrast to the previously mentioned works, the solutions proposed by~\cite{Eman,Kamil}, do not rely on a single entity for carrying out billing calculations. Instead, they involve a group of entities working together to compute individual bills/rewards privately. However, they are vulnerable to the risk of internal collusion, potentially leading to the disclosure of highly sensitive data such as individuals' meter readings and bid volumes. Additionally, market participants were uniformly charged for their deviations in all of the aforementioned works, irrespective of their positions on the grid, and the grid usage cost was not considered.


To address these limitations, in this paper, we propose a privacy-preserving billing protocol, considering participants' deviations and their locations on the grid. This work is an improvement on our previous study~\cite{Eman} to mitigate the impact of internal collusion on individuals' privacy. Specifically, the contributions of this paper are threefold:

\begin{itemize}
\item We propose an efficient and privacy-preserving individual billing scheme (EPIBS) that achieves an information theoretic security. EPIBS forms a building block for PBP-LEM. 
\item We design PBP-LEM -- a privacy-preserving billing protocol that enables suppliers to obtain bills for their contracted customers in LEMs. PBP-LEM consider participants' energy volume deviations and their locations on the grid in computing the bills without revealing any individuals private data. It involves a collaboration of different entities performing bill computation with verification of correctness. It also mitigates the potential impact on individuals' privacy resulting from internal collusion. It utilises EPIBS as well as other cryptographic primitives, namely multiparty computation (MPC), Pedersen commitments, and function-hiding inner product encryption (FHIPE), to compute individual bills. We propose three different approaches, resulting in different levels of privacy protection and performance.
 \item We evaluate the computation and communication complexity of PBP-LEM under honest-majority and dishonest-majority MPC. Our analysis shows performance variations and major overhead areas, depending on the employed approach. It also demonstrates the feasibility of PBP-LEM in real-world settings, with bill computation for 4,000 users taking less than five minutes even with the most computationally expensive approach. The implementation is publicly available at: \url{https://github.com/PBPLEMs/ZPPB-LEM2}.
\end{itemize}



 The rest of the paper is organised as follows. Section~\ref{sec:related_work} covers the related work. Section~\ref{sec:system-model} introduces our system model and its requirements. Section~\ref{sec:building-blocks} explains the utilised building blocks. Section~\ref{sec:protocol} describes PBP-LEM. Section~\ref{sec:analysis} provides security analysis, while Section~\ref{sec:evaluation} evaluates the performance of PBP-LEM. Finally, Section~\ref{sec:conclusion} concludes the paper.

\section{Related Work}\label{sec:related_work}
Concerns regarding security and privacy in LEMs have been previously highlighted~\cite{Mustafa-LEM}, prompting the proposal of several solutions using anonymisation, perturbation, and homomorphic encryption techniques. Only a few of these solutions have considered applying their privacy approach to a billing mechanism based on the actual amount of exported/imported energy by market participants, as recorded by their smart meters (SMs)~\cite{Xiaoyan,Gai,Gaybullaev,Son,Kamil,Andrei}. The work in~\cite{Kamil, Andrei} have also considered the participants' deviations from bid commitments.

The authors of~\cite{Gai,Xiaoyan} proposed a dynamic one-to-many accounts mapping technique to hide participant's trading distribution when their transactions are being recorded on blockchain. A key part of their works is a trusted entity, token bank, responsible for assigning each participant to multiple accounts based on a dynamically-styled bound that achieves the effect of deferential privacy. A major drawback of applying both~\cite{Gai} and~\cite{Xiaoyan} works in the real world  is the reliance on trusted identity management having access to user's accounts mappings and their transactions (i.e., which party could take this role in the real world and whether users are willing to reveal their transactions to them).

The works of~\cite{Gaybullaev} and~\cite{Son} mainly focused on protecting the privacy of market participants during the market clearance phase. The employed FHIPE enabling a market operator to execute the market and perform peer matching on encrypted bids without accessing their details. To hide individuals' participation in the market, participants (both sellers and buyers) are anonymised, and a new identifier (pseudonym) is generated for every new session to prevent linkability. However, after the market is executed, the distribution system operator takes on the responsibility of invoicing participants. In doing so, it gains access to the bid contents, and the correspondence between real identities and the one-time identifiers utilised in each session. This results in a similar limitation to that found in the works of ~\cite{Xiaoyan,Gai}, where there is a reliance on a party to handle the billing process, revealing both the real identities and bid details to that party.

The studies conducted by~\cite{Kamil,Andrei,Eman} are the only ones that have considered imperfect fulfilment of bid volumes. The work in~\cite{Andrei} proposed a privacy-preserving protocol for the three billing models outlined in~\cite{Madhusudan} by utilizing the Paillier homomorphic encryption technique. This protocol enables a trading platform (market operator) to compute individuals' bills/rewards without revealing their most sensitive data. In detail, during each trading period, each SM encrypts its meter reading and the deviation from its bid volume using the homomorphic public key of their contracted suppliers. Subsequently, it sends the encrypted data, along with other less sensitive information (e.g., the type of individuals' participation), to the trading platform. The platform then calculates individuals' bills privately in accordance with one of the billing models and sends the results to the suppliers, which in turn decrypts them. However, this work has certain shortcomings. Firstly, some of the participants' private data, such as the type of participation (i.e., whether a user is selling or buying energy) and whether the bid was accepted or rejected, is disclosed to the trading platform. Secondly, the protocol requires functional trust in a single party -- the trading platform -- for accurate computation of the participants' bills/rewards without any verification of correctness.

Similar to the approach taken by~\cite{Andrei}, the work in~\cite{Kamil} used the Paillier homomorphic encryption technique to implement a privacy-preserving billing model, accounting for participants' deviations from their commitments. However, in contrast to~\cite{Andrei}, this solution does not rely on a single party for executing the billing calculations. Instead, a randomly chosen set of market participants collaboratively computes individual bills/rewards in a private manner as follows: Individual private data are homomorphically encrypted using the supplier's keys and then transmitted to the chosen set, which, in turn, performs the computation privately. The solution also incorporates a referee to verify the correctness of bills/rewards computation simultaneously and intervene to resolve any disputes. The referee then sends the resultant encrypted bills/rewards to the supplier for decryption. However, a considerable limitation of this work is that it does not provide protection against collusion. For instance, any individual in the chosen set can collude with the supplier by providing it with others' encrypted individual data, subsequently enabling the supplier to decrypt the data and compromise the individuals' privacy. Another limitation is similar to that found in~\cite{Andrei}, wherein the type of individual participation is disclosed. 


Our previous work~\cite{Eman} leveraged MPC technique to apply a privacy-preserving billing model considering participants' deviations from their bid volumes. Instead of relying on a single trusted party for computing the bills, our approach employs three parties to jointly compute individual bills while ensuring the confidentiality of all the market participants' data, 
including their types of participation. The solution has also been evaluated under malicious and dishonest majority settings, implying that the presence of at least one honest party is sufficient to detect errors during bill calculations, even if the other two parties were malicious. Nevertheless, even though the probability of collusion among all three parties is low, given that each is controlled by a different entity, 
any such occurrence could lead to the disclosure of highly sensitive data. 


Unlike the aforementioned works, PBP-LEM integrates the following features: (i) it charges for market participants who fail to fulfil their bid commitments, (ii) it incorporates the participants' locations on the grid to distribute deviation costs fairly and accounts for distribution network usage fees in the participants' bills, (iii) it protects all participants' private data including their types of participation, (iv) it does not rely on a single trusted party to compute individual bills, and (v) it mitigates the impact of potential collusion among internal parties by ensuring that only less sensitive data might be revealed in the event of such collusion.


\begin{figure}[t]
\centering
\includegraphics[width=0.85\columnwidth,trim=4 4 4 4,clip]{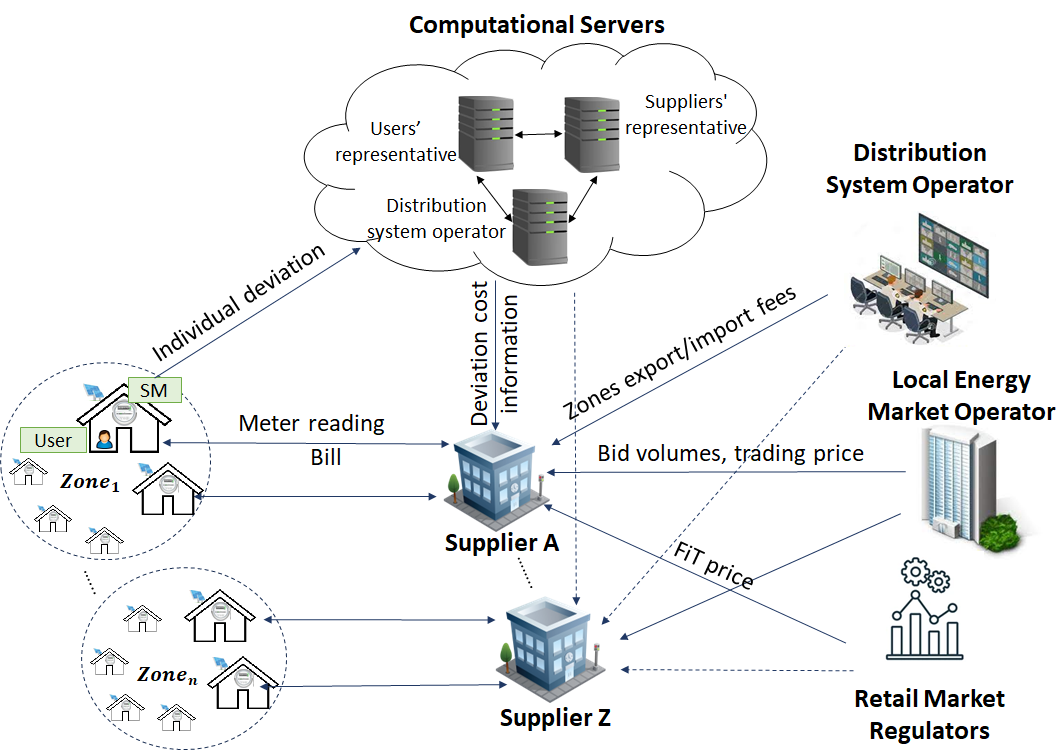}
  \captionsetup{justification=centering}
\caption{System model. }
\label{SystemModel}
\end{figure}
\section{System Model and Requirements}\label{sec:system-model}
In this section, we describe the system model, threat model as well as the functional and privacy requirements that PBP-LEM needs to satisfy.
\subsection{System Model}
As shown in Fig.~\ref{SystemModel}, our system model consists of the following entities. \textit{SMs} are devices that measure and communicate the volumes of imported and exported energy by households in nearly real-time. \textit{Users} are supported by SMs, in-home display device (connected to the SMs) and potentially potentially renewable energy sources. They engage through their in-home display devices in a LEM 
by the automatic submission of bids--comprising energy volume and offered price--to sell surplus energy to others at a price higher than FiT price or buy energy at a price lower than RP. \textit{Local Energy Market Operator (LEMO)} executes the LEM, identifying both the trading price and the approved bids for trading in each trading period. \textit{Suppliers} provide energy to users in need. They buy electricity from the wholesale market and distribute it to their contracted customers in the retail market at RP. They also purchase electricity injected by their customers that is not traded in the LEM at FiT price. Additionally, they generate and manage their customers' (i., users) bills based on their participation in the LEM. \textit{Retail Market Regulators (RMRs)} set FiT prices.\textit{Distribution System Operator (DSO)} manages and maintains the distribution network of a particular area. It divides the LEM area into small zones based on the physical network specifications and historical data that estimate each zone state for each time period. DSO also sets energy importing and exporting fees for each zone. It also coordinates the redistribution of payments to/from the LEM between users performed through their suppliers. \textit{Computational Severs (CS)} calculate the required deviation cost information (e.g., total global deviation) according to the zone-based billing model with universal cost split (ZBUCS) introduced in~\cite{Eman} which is a modification of the billing model proposed in~\cite{Madhusudan}. To avoid collusion, the parties should have conflicting interests. We assume that one party is controlled by suppliers, one by users, and a third by the DSO.



\subsection{Threat Model and Assumptions}
LEMO, RMRs and suppliers are assumed to be semi-honest. They follow PBP-LEM honestly, but they may try to infer sensitive individual data. Suppliers may attempt to learn individual bids, meter readings, rewards/bills within the LEM for a particular trading period, or may try to access the bills of users who are contracted with other suppliers. In addition, the correctness of bill settlements performed by suppliers must be verified by DSO. Users are considered to be malicious. They may try to learn or manipulate their own (or other users') data (i.e., bid volumes, meter readings and deviations) to increase their profits or reduce their bills. External entities are also malicious. They may eavesdrop or modify transmitted data to disturb the billing process in LEM. 

We assume two different security settings for CS: honest majority and dishonest majority with active (malicious) adversaries. For the former, one of the three parties can be corrupted while the later allows the presence of two malicious parties. The corrupted parties may try to infer information about users' data (bids, meter readings, bills and deviations) or deviate from the protocol, for example, by sending faulty data during the computations to distort the result. In the case of deviation, honest parties should abort the execution. 

As the focus of this work is on protecting users' privacy, we make the following assumptions. The communication channels are private and authentic. All entities are time-synchronized. SMs are tamper-proof and sealed, meaning any attempt to tamper with them will be readily detected.

\subsection{Functional Requirements}

\begin{itemize}

    \item Each supplier should learn each of their customers' aggregated bills/rewards generated from their participation in LEM per billing period (e.g., monthly). 
    \item Each supplier should learn its income balance incurred from selling (buying) the deviation shares to (from) their customers per trading slot. 
    \item Each user should learn their own LEM bill/reward per billing period. 
\end{itemize}
\subsection{Security and Privacy Requirements}
PBP-LEM should satisfy the following requirements:
  \begin{itemize}
       \item Confidentiality: users' bid volumes, meter readings, deviations, type of participation (selling or buying), bills/rewards per trading period from LEM as well as exact locations should be hidden from all parties.  
      \item Authorisation: aggregated users' bills and rewards from participating in LEM should be accessed only by their contracted suppliers.
      \item Collusion Impact Mitigation: users' highly sensitive data (i.e., bid volumes and meter readings) should not be revealed in the event of internal collusion.
      \item Accountability: reported capital exchanged within the LEM through each supplier should be verified for accuracy to ensure correct bill settlement.   
  \end{itemize}

\section{Building Blocks}\label{sec:building-blocks}
This section describes the foundational cryptographic techniques and building blocks utilized by PBP-LEM: function-hiding inner product functional encryption, pedersen commitment and multiparty computation. 

\subsection{Multiparty Computation}\label{Sec:MPC}
Multiparty Computation (MPC) enables a group of parties to jointly compute a function on their private inputs, revealing only the computation results and keeping individual data confidential~\cite{Lindell}. This can be achieved through various cryptographic techniques such as secret sharing, oblivious transfer, and homomorphic encryption. The security provided by these techniques can either be perfect, offering unconditional security against adversaries, or computational/conditional, ensuring security given computational limitations on adversaries. Information-theoretic protocols such as BGW provide perfect security~\cite{Ben-Or}, while protocols relying on public key primitives such as garbled circuits~\cite{Yao} and SPDZ~\cite{Ivan} offer conditional security.

An essential property of MPC protocols is how many parties can be corrupted. While information-theoretic protocols offer stronger (perfect) security, they require an honest majority~\cite{Lindell}. In contrast, computationally secure protocols can tolerate a dishonest majority; however, they tend to be more complex and expensive~\cite{Hastings}. 
The MPC protocols employed in our work for each security setting are as follows:

\begin{itemize}
\item \textit{Honest-Majority Setting}: In this setting, we utilize an optimized secret sharing approach from~\cite{Araki} based on replicated secret sharing, specifically designed for three parties. This protocol minimizes communication and computation costs as every party sends only one element for each multiplication gate using pseudo-random zero sharing. To achieve malicious security, we use the protocol proposed by~\cite{Chida}, which uses information-theoretic tags as in SPDZ protocol to check for computation correctness. The protocol supports security with abort, aiming for high efficiency.
\item \textit{Dishonest-Majority Setting}:
In this setting, we adopt the MASCOT protocol~\cite{Keller} for the malicious model. It is an improvement of the original SPDZ protocol, replacing the costly somewhat homomorphic encryption used for computing Beaver triples with oblivious transfer.

\end{itemize}

\subsection{Pedersen Commitments}
One of the core techniques used by PBP-LEM is Pedersen Commitments. In essence, it allows one party to commit to a secret value and later reveal the committed value in such a way that it becomes hard to change the initially committed value~\cite{Pedersen}. Based on the discrete logarithmic problem, Pedersen Commitments involve two fundamental functions:
\begin{itemize}
\item $c \leftarrow Commit(m,r)$: takes a value m and a random value $r$. It generates a commitment $c$. Mathematically, let $g$ and $h$ be generators of a cyclic group $G$, then $c= g^m . h^r mod p$.
\item $ true/false \leftarrow Open(c,m,r)$: takes a commitment $c$, a value $m$ and a random value $r$. It generates $true$ if $c$ is a commitment of $m$ and $false$ otherwise. 
\end{itemize}
Pedersen Commitment has three main properties. It ensures hiding, where given only a commitment $c$, it is computationally infeasible to determine the original value $m$. It upholds binding; that is, given $c$, $m$ and $r$, it is computationally infeasible to find a value $\bar{m}\neq m$ and a matching value $\bar{r}$ such that $Open(c,\bar{m},\bar{r})$ returns true. Finally, it satisfies a useful property known as homorphism, in which the product of two commitments equals the commitment to the sum of two initially committed values. 
\begin{equation*} Commit(m_1,r_1) * Commit(m_2,r_2) = Commit(m_1+m_2,r_1+r_2) \end{equation*}

\subsection{Inner Product Functional Encryption}
Functional encryption (FE) is a special type of public key cryptography that allows the performance of some computations over encrypted data. In FE schemes, the owner of the master key (i.e., key distribution centre) can generate secret keys for parties wishing to learn only outputs of specific functions over secret inputs and nothing else~\cite{Boneh}. In formal terms, given a ciphertext $C$ corresponding to a plaintext $x$ and a functional decryption key $sk$ related to a function $f$, a decryptor can obtain the results
of $f(x)$ without learning anything about $x$. 

\paragraph{Inner product functional encryption (IPE)} is a a special scheme of FE, where data $x$ is represented as a vector, functional decryption key $sk_y$ is associated to a vector $y$. Then, given the ciphertext $Ct_x$ of $x$, the owner of $sk_y$ can evaluate $<x.y>$ without learning $x$, which is the inner product of the vectors $x$ and $y$. FHIPE provides an additional property in which both $x$ and $y$ are kept secret from the function evaluator (i.e., the decryptor). In this work, we adopt the practical FHIPE scheme proposed in~\cite{Kim}. The original formulation of IPE schemes involves four algorithms: Setup, KeyGen, Encrypt, and Decrypt~\cite{Kim,Bishop}. However, for some cases, a more intuitive representation suggests KeyGen be referred to as LeftEncrypt and Encrypt as RightEncrypt, as discussed in~\cite{Kim}. The scheme involves the following four algorithms (for technical details, please refer to~\cite{Kim}):
\begin{itemize}
\item $(pp,msk) \leftarrow IPE.Setup(1_\lambda,S)$ given a security parameter $\lambda$, Setup initializes the system and outputs the public parameters $pp$ and  master key $msk$.
\item $Ct_x \leftarrow IPE.LeftEncrypt(msk,\alpha,x)$ given the master key $msk$, a vector $x$ and a uniformly random element $\alpha $, LeftEncrypt outputs a ciphertext $Ct_x$.
\item $Ct_y \leftarrow IPE.RightEncrypt(msk,\beta,y)$ given the master key $msk$, a vector $y$ and a uniformly distributed element $\beta$, RightEncrypt outputs $Ct_y$. 
\item $z \leftarrow IPE.Decrypt(pp,E_l(x),E_r(y))$ given $pp$, a ciphertext $Ct_x$. and a ciphertext $Ct_y$,  Decrypt outputs a message z such that $z=<x,y>$.
\end{itemize}

\subsection{Dual Binary Encoding for Integer Comparison Using Inner Products} 

In this work, we utilise a dual binary encoding scheme proposed by~\cite{Gaybullaev}. The scheme converts two integer numbers, $x$ and $y$, into arrays of vectors with a limited number of elements to enable comparing these two numbers by means of multiple inner products. More specifically, $x$ and $y$ undergo two different encoding methods:
\begin{itemize}
    \item $(X_{v_l}, X_{v_g}) \leftarrow f_x(x)$: on input a number $x$, the method $f_x$ output two encoded arrays of vectors $X_{v_l}=[x_{0,{l}},x_{1,l},...,x_{N,l}]$ and $X_{v_g}= [x_{0,g},x_{1,g},...,x_{N,g}]$.
    \item $ Y_v  \leftarrow f_y(y)$: on input a number $y$, the method $f_y$ output one encoded array of vector $Y_v = [y_0,y_1,...,y_N]$. \footnote{For technical details about the encoding processes, please refer to the work of~\cite{Gaybullaev}} 
\end{itemize}
By performing the inner products operations of $<X_{v_l},Y_v>$ and $<X_{v_g},Y_v>$ simultaneously, the results are interpreted as follows: 
\begin{itemize}
    \item If $<x_{n,l},y_n> = 0$ is satisfied first, it indicates that $x<y$.
 \item  If $<x_{n,g},y_n> = 0$ is satisfied first, it indicates that $y<x$.
 \item  If none of the above conditions is satisfied, this means $x=y$. 
\end{itemize}

\section{Privacy-Preserving Billing Protocol for Local Energy Market}\label{sec:protocol}
In this section, we propose PBP-LEM, considering the deviations cost. PBP-LEM is designed based on the billing model with a universal deviation cost split proposed in~\cite{Madhusudan}, which was later modified in our previous work~\cite{Eman} to incorporate users' locations on the grid. The main idea revolves around having the retail market as a backup market to compensate for users' deviations, which could come at a cost. The zone-based billing model with universal cost split, ZBUCS, introduced in~\cite{Eman} splits the total deviation cost among market participants whose deviations are in the same direction as the total deviation while considering the zones to which the participants belong on the grid. Prosumers (consumers) then pay (get paid by) their contracted suppliers for their total deviation share according to FiT (RP) rather than the LEM price, which is less beneficial. Please refer to the work of~\cite{Madhusudan} and ~\cite{Eman} for more details. 

It is important to note that this research specifically focuses on the billing phase of the LEM. The issue of preserving users' privacy during the market clearance phase has been addressed in earlier research, as referenced in~\cite{Alqahtani}. Our assumption is that user privacy during LEMO's market execution is protected by the adoption of privacy-preserving solutions outlined in previous studies, such as~\cite{Gaybullaev,Son,Abidin}. Consequently, the transmission of data by LEMO in PBP-LEM does not imply unrestricted access to individual user data by the entity. Instead, it only requires forwarding the securely protected data (e.g., bid volumes) to the computing parties after the market is executed to compute individual bills/rewards.
 
The section begins by proposing an efficient and privacy-preserving individual billing technique (Section ~\ref{sec:OurScheme}), which in turn is utilized by PBP-LEM (Section~\ref{sec:ourProtocol}). Table~\ref{Notations} lists the notation used throughout the paper.
\begin{table}
  \centering
  \footnotesize
  \begin{threeparttable}
  \caption{Notations}
\label{Notations}
\begin{tabular}{p{0.27 \columnwidth} p{0.73 \columnwidth}} 
\toprule
    \textbf{Symbol} & \textbf{Description} \\
       \midrule
       $bp_u$ & $u$-th billing period, $u \in \{1, ..., N_u\}$. \\
    $tp_k$ & $k$-th trading period, $k \in \{1, ..., N_k\}$. \\ 
     $ID_i$ & Unique identifier of user $i, i \in \{1, ..., N_i\}$. \\
    $sID_j, zID_z$ & Unique identifier of of user $i$, supplier $j$ and zone $z$.\\
       $b_i^{tp_k}, m_i^{tp_k}, v_i^{tp_k}$, $d_i^{tp_k}$, $s_i^{tp_k}$, $bl_i^{tp_k}$ & Bid volume, meter reading, individual deviation, type of participation (binary value, 1 for producer and 0 for consumer), deviation cost inclusion state (binary value), bill/reward
 of user $i$ at $tp_k$.  \\
      $BL_i^{bp_u}$ & Bill/reward of user $i$ at $bp_u$.  \\
      $BL_i^{{bp_u},LEM}$ & portion of bill/reward dedicated to LEM of user $i$ at $bp_u$.  \\
      $SBal_j^{tp_k}$ & Income balance of supplier $j$ resulting from users trading of deviations cost at $tp_k$. \\
       $SCap_j^{bp_u}$ & capital traded exclusively within the LEM through supplier $j$ at ${bp_u}$.\\
      $pp_i,msk_i$ & Public parameters and master key of FHIPE for user $i$. 
      \\
      $S_i, S_i^t$ & Two set of random keys for EPIBS corresponding to the trading periods $\{tp_1, \ldots, tp_{N_k}\}$ within $bp_u$ generated for user $i$\\ 
        $C_{i,p}^{tp_k}$, $C_{i,c}^{tp_k}$ &  Condition that requires user $i$ to pay/get paid to/by supplier due to his/her deviation at $tp_k$. \\
      $mc_i^{tp_k},dc_i^{tp_k}, bc_{i}^{tp_k}$ & Ciphertexts of $m_i^{tp_k},d_i,^{tp_k}$ with $sk_i^{tp_k},sk_{i,t}^{tp_k}$, respectively, ciphertext of $bl_i^{tp_k}$, at $tp_k$ using EPIBS.\\
       $DK_i^{bp_u}$ & Decryption key of $\sum_{k=1}^{N_k} bc_i^{tp_k}$ using EPIBS. \\
      $Mc_i^{tp_k}$, $(Bc_{i,v_l}^{tp_k} , Bc_{i,v_g}^{tp_k})$ & Ciphertexts of dual bianry encoded $m_i^{tp_k}$, $b_i^{tp_k}$ at $tp_k$ with  $msk_i$ using $IPE.LeftEncrypt$ and $IPE.RightEncrypt$, respectively. \\
        $t_z^{tp_k},np_z^{tp_k}/nc_z^{tp_k}$ & Total deviation, number of producers/consumers in zone $z$ at $tp_k$\\
              $T^{tp_k}, W^{tp_k}$ & Total LEM deviation, zonal deviation weight at $tp_k$. \\[1.05ex]
           $TP^{tp_k}, FiT^{tp_k}, RP^{tp_k}$ & LEM trading price, FiT and RP at $tp_k$.  \\

\bottomrule

\end{tabular}
\end{threeparttable}
\end{table}
\subsection{Efficient and Privacy-Preserving Individual Billing Scheme}\label{sec:OurScheme}
In this section, we propose an efficient and privacy-preserving individual billing scheme, EPIB, that achieves information-theoretic security and forms a building block for PBP-LEM. 
Initially, let's assume that a set of encryption keys $S_i = \{sk_i^{tp_1},\ldots,sk_i^{tp_{N_k}}\}$ is generated for every user, corresponding to trading periods $\{tp_1, \ldots, tp_{N_k}\}$ within one billing period, $bp_u$. Each encryption key $sk_i^{tp_k}$ is a randomly generated number over a finite field modulo q $(z_q)$. Then, our privacy-preserving individual billing technique operates over $z_q$ as follows. 

After each trading period, $tp_{k}$, each user, $i$, encrypts its recorded meter reading, $m_i$, during $tp_{k}$ using the corresponding $sk_i^{tp_k}$ by simply computing: 
\begin{equation*}
mc_i^{tp_k} = m_i^{tp_k} + sk_i^{tp_k} 
\end{equation*} \label{eq:Encrypton}

Given the identified market trading price, $TP^{tp_k}$, to compute the encrypted individual bill/reward, one performs the following computation:

\begin{equation*} bc_i^{tp_k} = mc_i^{tp_k} * TP^{tp_k}\end{equation*} \label{eq:BillComputaion}

Including the deviation cost in users' bills/rewards based on ZBUCS requires knowledge of the user's type of participation, $d_i^{tp_k}$ -- that is, whether the user offered to produce or consume energy during ${tp_k}$. To protect $d_i^{tp_k}$, let $S_i^t= \{sk_{i,t}^{tp_1}, \ldots, sk_{i,t}^{tp_{N_k}}\}$ represents a set of randomly generated keys over $z_k$, corresponding to the trading periods $\{tp_1, \ldots, tp_{N_k}\}$ within $bp_u$. For each $tp_k$, each user masks $d_i^{tp_k}$ using $sk_{i,t}^{tp_k}$.
\begin{equation*}
dc_i^{tp_k} = d_i^{tp_k} + sk_{i,t}^{tp_k}
\end{equation*} \label{eq:PMasking}

Let $dev_p^{tp_k}$ denote the cost to be paid by certain producers to their suppliers due to deviations based on a condition, $C_{i,p}^{tp_k}p$, and $dev_c^{tp_k}$ denote the compensation to be received by specific consumers from their suppliers for deviations based on a condition, $C_{i,c}^{tp_k}$, as defined in ZBUCS ($C$ is equal to 1 if the condition is true, and 0 otherwise). Then, for each $tp_k$, the addition of $dev_p^{tp_k}$ and $dev_c^{tp_k}$ to $bc_i^{tp_k}$ is as follows:
\begin{multline*}
bc_i^{tp_k} = bc_i^{tp_k} + C_{i,p}^{tp_k}(dc_i^{tp_k} * dev_p^{tp_k}) + C_{i,c}^{tp_k}((1 - dc_i^{tp_k}) * dev_c^{tp_k})
\end{multline*}

The decryption key for ${tp_k}$ is then computed as:

\begin{equation*} dk_i^{tp_k} = (sk_i^{tp_k} * TP^{tp_k}) + C_{i,p}^{tp_k}(sk_{i,t}^{tp_k} * dev_p^{tp_k}) - C_{i,c}^{tp_k}( sk_{i,t}^{tp_k} * dev_c^{tp_k})
\end{equation*} \label{eq:decKey}

For every $bp_u$, the encrypted monthly bill, $BLc_i^{bp_u}$, is calculated by summing the results $bc_i^{tp_k}$ of each ${tp_k}$: $BLc_i^{bp_u} = \sum_{k=1}^{N_k} bc_i^{tp_k}$. Similarly, the decryption key $DK_i^{bp_u}$ for $bp_u$ is computed by summing the results $dk_i^{tp_k}$ for every ${tp_k}$: $DK_i^{bp_u} = \sum_{k=1}^{N_k} dk_i^{tp_k}$. Finally, to decrypt, one simply computes: 

\begin{equation*} BL_i^{bp_u} =  BLc_i^{bp_u} - DK_i^{bp_u} \end{equation*} \label{eq:decryptionKey}

The above scheme guarantees the protection of $m_i^{tp_k}$ and $d_i^{tp_k}$. Its security comes from the fact that if $sk_i^{tp_k}$ and $sk_{i,t}^{tp_k}$ are fresh random numbers generated for every $tp_k$, then the scheme can be seen as equivalent to one-time pad.  

\subsection{Privacy-Preserving Billing Protocol}\label{sec:ourProtocol}
This section provides a detailed description of PBP-LEM considering the deviations cost. PBP-LEM consists of five steps: input data generation and distribution, bills computation per trading period, bills computation and distribution per billing period, individual deviations verification, and bills settlement and verification. We first discuss some prerequisite steps that should be performed. We then discuss the five protocol steps in detail, with general overviews provided in Fig.~\ref{OverviewPT} and Fig.~\ref{OverviewPB} in Appendix~\ref{App:Overview}. The discussion includes a description of three different approaches applied in PBP-LEM, leading to varying degrees of privacy and performance. The square brackets [x] denote that x is secretly shared, and the angles $\langle x \rangle$ denote a Pedersen commitment to x. 


\subsubsection{Prerequisites}
Before participating in LEM, users’ keys are generated. These keys can be generated either by the users’ SMs (with proof of correct correspondence between the encryption and decryption keys, e.g., using Zero-Knowledge Proofs) or the task could be delegated to a key authority (KA), which could be operated by a regulatory authority such as the Office of Gas and Electricity Markets in the UK. In the latter case, each user would send a registration request along with their unique identifier, $ID_i$, their contracted supplier identifier, $sID_j$, to the KA. Subsequently, Subsequently, KA generates two sets of random keys, following EPIBS. It also generates the master key and public parameters of the FHIPE scheme~\cite{Kim}, $(pp_i,msk_i) = IPE.Setup(1_\lambda,S)$. KA then transmits the tuple \{$S_i, S^t_i,msk_i$\} to the user and its SM and the tuple \{$ID_i, pp_i$\} to its corresponding supplier. For every billing period $bp_u$, new sets of keys $S_i$ and $S^t_i$ are generated for each user. 


Additionally, we make the following assumptions: bid volumes, $b_i^{tp_k}$, are encoded using  $f_x$ of the encoding scheme ~\cite{Gaybullaev},$(B_{i,{vl}}^{tp_k},B_{i,{vg}}^{tp_k}) = f_x(b_i^{tp_k})$, and then encrypted it with $msk_i$ using FHIPE, 
$Bc_{i,{vl}}^{tp_k} = (IPE.RightEncrypt(msk_i,\beta,B_{i,{vl},j}^{tp_k})$ for all vectors of $B_{i,{vl}}^{tp_k}$), $Bc_{i,{vg}}^{tp_k} = (IPE.RightEncrypt(msk_i,\beta,B_{i,{vg},j}^{tp_k})$ for all vectors of $B_{i,{vg}}^{tp_k}$).  LEMO receives $Bc_{i,{vl}}^{tp_k}$ and $Bc_{i,{vg}}^{tp_k}$ from users before the market clearance phase to execute the market privately, as in~\cite{Gaybullaev}. Please note that the above assumption is required only for approach 1 in step 2. We also assume that each user sends $dc_{i,t}^{tp_k}$ and a Pedersen commitment to the negative inverse of their bid volumes, $\langle -b_i^{tp_k} \rangle = Commit(-b_i^{tp_k},r_{i,b}^{tp_k})$, to LEMO along with their submitted bids before the market execution. Next, we discuss the five steps of PBP-LEM.

\subsubsection{Step 1: Input Data Generation and Distribution}
After each trading period $tp_k$, tuples are prepared by users, their respective SMs, and LEMO. These tuples are then shared with CS and suppliers, enabling collaborative computation of the users' bills and rewards from the LEM, inclusive of their deviations cost. In detail, every user in every zone calculates their deviation $v_i^{z,tp_k}$ by subtracting $m_i^{tp_k}$ from $b_i^{tp_k}$. Each user then generates a tuple \{$ID_i,sID_j,zID_z,[v]_i^{tp_k},[d]_i^{tp_k}$\} which contains the zone to which the user belongs, $zID_z$, shares of the computed deviations, $[v]_i^{tp_k}$ and shares of their type of participation $[d]_i^{tp_k}$ (i.e., a producer or consumer) and sends it to the CS. 

 \begin{algorithm}[t]
      \footnotesize
        \caption{Zone-based Deviations Aggregation}\label{alg:DeviationsAggr}
        \hspace*{\algorithmicindent} \textbf{Input:} Set of $N_i^z$ user tuples $U = ([v],[d]$) who belong to zone z \\
        \hspace*{\algorithmicindent} \textbf{Output:} Zone $z$ tuple $ZN= ([t],[np],[nc]$)
        \begin{algorithmic}
        \For{$i = 0$ to $N_i^z$ }
            \State $[t]_z^{tp_k} \gets [t]_z^{tp_k} + [v]_i^{tp_k}$
            \State $[np]_z^{tp_k} \gets [np]_z^{tp_k} + [d]_i^{tp_k}$
            \State $[nc]_z^{tp_k} \gets [nc]_z^{tp_k} + 1-[d]_i^{tp_k}$
        \EndFor
        \end{algorithmic}
        \end{algorithm}

In addition, every SM encodes $m_i^{tp_k}$ using $f_y$ of the scheme ~\cite{Gaybullaev}, $M_i^{tp_k} = f_y(m_i^{tp_k})$, and then encrypts it with $msk_i$ using FHIPE,  $Mc_i^{tp_k} = (IPE.LeftEncrypt(msk_i,\alpha, M_{i,j}^{tp_k})$ for all vectors of $M_i^{tp_k}$). This operation is required by only Approach 1 in Step 2. The SM also encrypts $m_i^{tp_k}$ with $sk_i^{tp_k}$ using EPIB, $mc_i^{tp_k}$. It also commits to $m_i^{tp_k}$, $\langle m_i^{tp_k}\rangle = Commit(m_i^{tp_k},r_{i,m}^{tp_k})$. Upon completion, the SM constructs the tuple \{$ID_i, zID_z, mc_i^{tp_k},Mc_i^{tp_k}$, $\langle m_i^{tp_k}\rangle$\} and transmits it to the designated supplier. Additionally, LEMO forwards the following tuple \{$ID_i,sID_j$, $Bc_{i,v_l}^{tp_k},Bc_{i,v_g}^{tp_k},\langle -b_i^{tp_k} \rangle,dc_i^{tp_k}$\} for each user to their suppliers. 

\subsubsection{Step 2: Bills Computation Per Trading Period}\label{step2}
Upon the receipt of the tuples, firstly, CS jointly computes what is required to include users' deviations cost in their bills/rewards. The CS then sends the results to the suppliers, which subsequently continue the computations to calculate users' total bills/rewards per $tp_k$  in a data oblivious fashion. In detail, once the CS receives \{$ID_i,sID_j,zID_z,[v]_i^{tp_k},[d]_i^{tp_k}$\} from users, they evaluate the total deviation, $[t_z]^{tp_k}$, number of consumers, $[nc_z^{tp_k}]$, and number of prosumers, $[np_z]^{tp_k}$, per zone obliviously as shown in algorithm~\ref{alg:DeviationsAggr}. The CS then reconstruct $t_z^{tp_k}$, $nc_z^{tp_k}$ and $np_z^{tp_k}$ for each zone using their shares in order to compute the total global deviation, $T^{tp_k}$, and the zonal deviation weight, $W^{tp_k}$, which help to distribute the total global deviation proportionally between the zones (For more details about the calculation of $W^{tp_k}$, please refer to~\cite{Eman}). Once completed, the CS, publish the set of $N_z$ tuples $ZN =  (t_z^{tp_k}, nc_z^{tp_k}, np_z^{tp_k})$, resulting from algorithm~\ref{alg:DeviationsAggr}, as well as ($T^{tp_k}, W^{tp_k}$) tuple to all suppliers, users and KA.

\begin{algorithm}[t]
      \footnotesize
        \caption{$s_i^{tp_k}$ identification by suppliers (\textbf{Approach 1})}\label{alg:IPEComparision}
        \hspace*{\algorithmicindent} \textbf{Input:}  $Mc_{i}^{tp_k}$, $Bc_{i,v_l}^{tp_k},Bc_{i,v_g}^{tp_k} T^{tp_k},t_z^{tp_k}$ to which the user belongs. \\
        \hspace*{\algorithmicindent} \textbf{Output:} $s_i^{tp_k}$
        \begin{algorithmic}
        \State $s_i^{tp_k} \gets 0$ 
         \For{$j = 0$ to $N_v$ }
                      \If{$IPE.Decrypt(pp_i, Mc_{i,j}^{tp_k}, Bc_{i,v_l,j}^{tp_k}) = 0$}  \Comment{$b_i^{tp_k} < m_i^{tp_k} \rightarrow v_i^{tp_k}>0$}
                         \If{$T^{tp_k} >0$ and $t_z^{tp_k} >0$}
                             \State $s_i^{tp_k} \gets 1$    
                         \EndIf
                         \Return $s_i^{tp_k}$ 
                      \EndIf
                      \If{$IPE.Decrypt(pp_i, Mc_{i,j}^{tp_k}$, $Bc_{i,v_g,j}^{tp_k}) = 0$} 
                        \If{$T^{tp_k} <0$ and $t_z^{tp_k} <0$}
                           \State $s_i^{tp_k} \gets 1$ 
                        \EndIf
                       \Return $s_i^{tp_k}$ \Comment{$b_i^{tp_k} > m_i^{tp_k} \rightarrow v_i^{tp_k}<0$}
                      \EndIf
                  
            \EndFor
        \Return $s_i^{tp_k}$ 
        \end{algorithmic}
        \end{algorithm}

Upon receiving the necessary data from the CS, suppliers can begin calculating LEM bills and rewards for the users, who are their contracted customers in the retail market. Initially, suppliers need to determine a binary value, $s_i^{tp_k}$, for each of their customers, indicating if the user must pay a deviation cost (the part of the user's production/consumption they have to sell (buy) to (from) their suppliers according to $FiT^{tp_k}$ or $RP^{tp_k}$ rather than $TP^{tp_k}$) based on ZBUCS~\cite{Eman}. $s_i^{tp_k}$ is equal to 1 if the user has to pay a deviation cost and 0 otherwise.

\subsubsection{Approach 1: $s_i^{tp_k}$ Identification Performed by Suppliers}
In this approach, each supplier identify $s_i^{tp_k}$ for their customers by themselves without the knowledge of $m_i^{tp_k}$ or $b_i^{tp_k}$. They utilise the encoding and comparison scheme of~\cite{Gaybullaev}; and FHIPE~\cite{Kim}. Specifically, for each user, using $Bc_{i,v_l}^{tp_k}$, $Bc_{i,v_g}^{tp_k}$ obtained from LEMO and $Mc_i^{tp_k}$ received from the SM, the supplier compares $m_i^{tp_k}$ and $b_i^{tp_k}$ values and determine $s_i^{tp_k}$ in an oblivious manner as shown in Algorithm~\ref{alg:IPEComparision}.

\subsubsection{Approach 2: $s_i^{tp_k}$ Identification by CS}
In this approach, the CS take charge of identifying $s_i^{tp_k}$ instead of leaving it to the suppliers. The CS continue to perform computations on $[v_i^{tp_k}]$ by executing comparisons as detailed in Algorithm~\ref{alg:SPCSComparision}. Subsequently, the CS send the resultant shares, $[s_i^{tp_k}]$, to the designated supplier according to $sID_j$, allowing them to recover $s_i^{tp_k}$.

 \begin{algorithm}[t]
      \footnotesize
        \caption{$s_i^{tp_k}$ identification by CS (\textbf{Approach 2})}\label{alg:SPCSComparision}
        \hspace*{\algorithmicindent} \textbf{Input:}  $T^{tp_k},t_z^{tp_k}$ to which the user belongs. \\
        \hspace*{\algorithmicindent} \textbf{Output:}  $[s]_i^{tp_k}$
        \begin{algorithmic}
        	\State $[s]_i^{tp_k} \gets 0 $
            \If{$T^{tp_k} >0$ and $t_z^{tp_k} >0$}
                \State $[s]_i^{tp_k} \gets [v]_i^{tp_k}>0$
            \ElsIf{$T^{tp_k} <0$ and $t_z^{tp_k} <0$}
                \State $[s]_i^{tp_k} \gets [v]_i^{tp_k}<0$
            \EndIf
        \end{algorithmic}
        \end{algorithm}
        
\subsubsection{Approach 3: Individual Deviations Disclosed to Suppliers}
Since individual deviations are less sensitive data, we allow their disclosure to suppliers, enabling them to compare the values in clear. While this disclosure would reveal some information about users, critical sensitive data such as meter readings and bid volumes should still not be inferred. This avoids the significant overhead associated with performing comparisons on encrypted or secretly shared data.

After identifying $s_i^{tp_k}$, suppliers proceed with the computation of users' bills/rewards per $tp_k$, using EPIBS as shown in Algorithm~\ref{alg:IndividualBilling}. Specifically, for each user, their respective supplier calculates $bc_i^{tp_k}$ based on $mc_i^{tp_k}$ received from the SMs, and $dc_i^{tp_k}$ obtained form LEMO (\textit{line 5}). In addition, the supplier updates its income balance resulting from users' deviations, denoted as $SBal_j^{tp_k}$, by adding revenue and subtracting expenditures incurred from selling (or buying) the deviations costs to (or from) users (\textit{line 6}). Upon completion, the supplier reports \{$ID_i, C_{i,c}^{tp_k}$, $C_{i,p}^{tp_k}$\} to KA for decryption key generation of EPIBS. KA then computes $dk_i^{tp_k}$ using $sk_i^{tp_k}$, $sk_{i,t}^{tp_k}$ based on EPIBS. It also generates the decryption key required for the supplier to obtain its income balance per $tp_k$, $DK_j^{tp_k} = \sum_{i=1}^{N_i^j} (-C_{i,p}^{tp_k}(sk_{i,t}^{tp_k} * dev_p^{tp_k}) + C_{i,c}^{tp_k}( sk_{i,t}^{tp_k} * dev_c^{tp_k}))$.

\begin{algorithm}[t]
      \footnotesize
        \caption{Individual Bill Computation Per Trading Period}\label{alg:IndividualBilling}
        \hspace*{\algorithmicindent} \textbf{Input:}  $mc_i^{tp_k}$, $dc_i^{tp_k}$, $SBal_j^{tp_k}$, ($T^{tp_k}, W^{tp_k}$), zone z tuple $ZN =  (t_z^{tp_k}, nc_z^{tp_k}, np_z^{tp_k})$ to which the user belongs.\\
        \hspace*{\algorithmicindent} \textbf{Output:} $bc_i^{tp_k}$, $SBal_j^{tp_k}$, $C_{i,p}^{tp_k}$ , $C_{i,c}^{tp_k}$
        \begin{algorithmic}[1]
        	 	\State $C_{i,p}^{tp_k} \gets (T^{tp_k} > 0 $  \textbf{and}  $s_i^{tp_k} )$ ; $C_{i,c}^{tp_k} \gets (T^{tp_k} < 0 $  \textbf{and}  $s_i^{tp_k})$
            \State $dev_p^{tp_k} \gets (t_z^{tp_k} \times W^{tp_k}/ np_z^{tp_k}) \times (FiT^{tp_k} - TP^{tp_k}) $
             \State $dev_c^{tp_k} \gets (t_z^{tp_k} \times W^{tp_k}/ nc_z^{tp_k}) \times (RP^{tp_k} - TP^{tp_k}) $ 
             \State $bc_i^{tp_k} \gets EPIBS.BillComp(mc_i^{tp_1}, dc_i^{tp_k}, C_{i,p}^{tp_k}, C_{i,c}^{tp_k})$
             \State $SBal_j^{tp_k} \gets SBal_j^{tp_k} -  C_{i,p}^{tp_k} \times dc_i^{tp_k} \times (t_z^{tp_k} \times W^{tp_k}/ np_z^{tp_k}) \times FiT^{tp_k}$
              \State $SBal_j^{tp_k} \gets SBal_j^{tp_k}  -  C_{i,c}^{tp_k} \times (1 - dc_i^{tp_k}) \times  (t_z^{tp_k} \times W^{tp_k}/ nc_z^{tp_k}) \times RP^{tp_k}$
        \end{algorithmic}
  \end{algorithm} 
\subsubsection{Step 3: Bills Computation and Distribution Per Billing Period}
At the end of the billing period $bp_u$, suppliers finalize the computation of users' bills/rewards using EPIBS. For each user, their corresponding supplier aggregates $bc_i^{tp_k}$ for all trading periods within $bp_u$: $BLc_i^{bp_u} = \sum_{k=1}^{N_k} bc_i^{tp_k}$. In addition, KA generates the decryption key for $bp_u$, $DK_i^{bp_u} = \sum_{k=1}^{N_k} dk_i^{tp_k}$ for each user and sends them to their suppliers according to $sID_j$.  Upon receiving $DK_i^{bp_u}$ from KA, the supplier decrypts $BLc_i^{bp_u}$ using EPIBS to obtain the user's bill/reward, $BL_i^{bp_u}$, which is then sent to the user. As each user has all the required information (e.g., $T^{tp_k}$ and $W^{tp_k}$) to compute its individual bill/reward, the user accepts $BL_i^{bp_u}$ only if it is identical to its locally computed bill. Additionally, for subsequent verification, each user calculates its bill/reward dedicated solely to the LEM excluding the deviation cost, $BL_i^{{bp_u},LEM}$(see Algorithm~\ref{alg:BillsComputation} in Appendix~\ref{App:BillsComputation}).

\subsubsection{Step 4: Individual Deviations Verification}
The purpose of this step is to verify the accuracy of the individual deviations $[v]_i^{tp_k}$ computed by each user and transmitted to the CS.
Before getting into details, it is important to note that this is an ongoing process conducted for every $tp_k$ and at the end of ${bp_u}$. Pedersen commitment is utilised to verify the accuracy of $[v]_i^{tp_k}$ for each user. For every $tp_k$, given $\langle m_i^{tp_k} \rangle$ received from an SM and $\langle -b_i^{tp_k} \rangle$ received from LEMO, the designated supplier exploits the homomorphism property of Pedersen commitment to obtain a commitment to $v_i^{tp_k}$:

$$ \langle v_i^{tp_k} \rangle = \langle m_i^{tp_k} \rangle . \langle -b_i^{tp_k} \rangle$$

At the end of ${bp_u}$, utilising the same homomorphism property, the supplier computes a commitment to the aggregated $v_i^{tp_k}$ during ${bp_u}$ through multiplication: 

$$\langle\sum_{k=1}^{N_k} v_i^{tp_k}\rangle = \prod_{k=1}^{N_k}\langle v_i^{tp_k} \rangle $$

Concurrently, the CS aggregates all $[v_i]^{tp_k}$ values during ${bp_u}$, for each user:
 $$[V_i^{bp_u}] = \sum_{k=1}^{N_k} [v_i^{tp_k}]$$ 
The CS then retrieve the result, $V_i^{bp_u}$, by reconstructing their shares and send it to the designated supplier according to $sID_j$. Additionally, each SM sums all the random numbers, $R_i^{bp_u} = \sum_{k=1}^{N_k} (r_{i,m}^{tp_k} + r_{i,b}^{tp_k})$, used earlier in its previous commitments (i.e., $ \langle m_i^{tp_k} \rangle$ and $\langle -b_i^{tp_k} \rangle$) and sends the total $R_i^{bp_u}$ to the designated supplier. Finally, the supplier checks that $V_i^{bp_u}$ is correct if opening the commitment $<\sum_{k=1}^{N_k} v_i^{tp_k}>$ using $V_i^{bp_u}$ and $R_i^{bp_u}$ returns true. 
$$Open(<\sum_{k=1}^{N_k} v_i^{tp_k}>,V_i^{bp_u},R_i^{bp_u}) \stackrel{?}{=}  true$$ This, in turn, indicates that all previous $v_i^{tp_k}$ shared by user $i$ with the CS were correct.

\subsubsection{Step 5: Bills Settlement and Verification}
In this final step, suppliers settle billing period bills with users who are their contracted customers, ensuring that any payments to or from the LEM, sent or received by these customers, are appropriately allocated among other suppliers. Specifically, each supplier conducts billing with their customers based on $BL_i^{bp_u}$, which encompasses both LEM payments/rewards and deviations share payments/rewards dedicated to the suppliers. The supplier, then, separates these two values. First, the supplier computes its accurate income balance for every $tp_k$ using $Dk_j^{{tp_k}}$ received from KA, $SBal_j^{tp_k} = SBal_j^{tp_k} - Dk_j^{{tp_k}}$. The supplier, then, subtract its total income balance during $bp_u$ from the total of its customers bills to compute the remaining capital exchanged within the LEM through it, $$SCap_j^{bp_u} = -\sum_{i=1}^{N_i^j} BL_i^{bp_u} - \sum_{k=1}^{N_k} SBal_j^{tp_k}$$

$SCap_j^{bp_u}$ values are redistributed among other suppliers through the coordination of the DSO. The sum of these values should be zero, indicating that all the LEM transactions have been accounted for legitimately. In the case of a non-zero sum, the DSO should intervene to verify the correctness of each $SCap_j^{bp_u}$ value. To do so, the DSO collects tuples \{$sID_j, BL_i^{{bp_u},LEM}$\} from all users. The DSO then segregates and aggregates $BL_i^{{bp_u},LEM}$ values based on their supplier correspondence $sID_j$, such that $SCap_j^{{bp_u},DSO}$ represents the leftover capital traded at the LEM and computed by the DSO. The DSO then validates every $SCap_j^{bp_u}$ value by comparing it with $SCap_j^{{bp_u},DSO}$. Any mismatch indicates that supplier $j$ has been dishonest in submitting an invalid $SCap_j^{bp_u}$ value. 

\section{Security Analysis}\label{sec:analysis}
We analyse the security of PBP-LEM under the universal composability (UC) framework~\cite{Canetti,Canetti2}. The framework provides a security gurantees by which a protocol proven secure in this framework is assured to maintain its security when it is run sequentially or concurrently with other protocols.  The framework consists of the real world (i.e. the real protocol), ideal world and an environment. In the ideal world, the parties do not communicate with each other. Instead they have access to an ideal functionality (i.e., “trusted party”) that is defined to capture the desired security requirements of the real protocol. Then, security of the real protocol is implied if no environment can computationally distinguish between the two worlds. 


\begin{definition}[UC emulation~\cite{Canetti}] A real protocol $\pi$ UC-realizes an ideal functionality $\mathcal{F}$ if for any real-world adversary $\mathcal{A}$, there exists a simulator $\mathcal{S}$ such that no environment $\mathcal{Z}$ is able to distinguish with a non-negligible probability between an interaction with $\mathcal{A}$ and real parties running protocol $\pi$ and an interaction with $\mathcal{S}$ and dummy parties accessing $\mathcal{F}$, i.e., $REAL_{\pi,\mathcal{A},\mathcal{Z}} \approx IDEAL_{\mathcal{F},\mathcal{S},\mathcal{Z}}$.
\end{definition}

\begin{definition}[Composition Theorem~\cite{Canetti}] 
Let $\rho$, $\varphi$, and $\pi$ be protocols such that $\varphi$ is a subroutine of $\pi$, $\rho$ UC-emulates $\varphi$, and $\pi$ UC-realizes an ideal functionality $\mathcal{F}$, then protocol $\pi^{\varphi \rightarrow \rho}$ UC-realizes $\mathcal{F}$.
\end{definition}

We assume the presence of the following ideal functionalities: secure computation ideal functionality $\mathcal{F}_{SecComp}$ for MPC and IPFE; and a commitment ideal functionality $\mathcal{F}_{COM}$.  $\mathcal{F}_{SecComp}$ receives private inputs from users, preform computation on the inputs and reveal the results upon parties agreement. $\mathcal{F}_{COM}$ receives values to be committed from users, stores them in its internal state and reveal them upon request.  We also construct an ideal functionality for EPIBS $\mathcal{F}_{EPIBS}$ as shown in Fig.~\ref{fig:FEPIBS}. For simplicity, the functionality solely captures the computation of individual bills with no inclusion of the deviations cost. 


\begin{theorem}
    Let $\pi_{\text{EPIBS}}$ be the billing protocol employing EPIBS, excluding the consideration of the deviations cost.  Then the protocol $\pi_{\text{EPIBS}}$ securely emulates the functionality $\mathcal{F}_{EPIBS}$.
\end{theorem}

\textit{PROOF.} We construct a simulator $\mathcal{S}$ such  that no environment $\mathcal{Z}$ can distinguish between an ideal and real execution. $\mathcal{S}$ executes a simulated copy of $\mathcal{A}$ and imitates an interaction with parties running $\pi_{\text{EPIBS}}$ to make the simulation of internal protocol interactions consistent with the real execution. 
$\mathcal{S}$ proceeds as follows: upon receiving $(ID_i,l(m_i^{tp_k}+sk_i^{tp_k}), l(sk_{i,t}^{tp_k}))$ from $\mathcal{F}_{EPIBS}$, $\mathcal{S}$ mimics the encryption operations of EPIBS on behalf of the smart meter. It defines $mc_i^{tp_k}$, $dc_i^{tp_k}$ to be random values $\Delta_{mc_i^{tp_k}}$, $\Delta_{dc_i^{tp_k}}$ from the appropriate space given $l(m_i^{tp_k}+sk_i^{tp_k}), l(sk_{i,t}^{tp_k})$. $\mathcal{S}$ then feeds $\mathcal{A}$ with the values $\Delta_{mc_i^{tp_k}}$, $\Delta_{dc_i^{tp_k}}$.
 
Now given that $sk_i^{tp_k}$ and $sk_{i,t}^{tp_k}$ are truly random numbers, it implies that $mc_i^{tp_k}$, and $dc_i^{tp_k}$ based on EPIBS, are uniformly distributed. This indicates perfect indistinguishably as it is impossible to distinguish between $mc_i^{tp_k}$ and $\Delta_{mc_i^{tp_k}}$ (since the distribution over the ciphertext is the same in every case as in one-time pad). Therefore , we can conclude that the output of the of the real world and ideal world executions are indistinguishable. 

\begin{figure}[t]
    \centering
    \footnotesize
    \begin{tikzpicture}
        \node [draw, text width=11.6cm, inner sep=10pt] (textbox) at (0,0) {
            \centerline{\textbf{Functionality $\mathcal{F}_{EPIBS}$}}
            $\mathcal{F}_{EPIBS}$ proceeds as follows, when parameterized by a mapping between users and their suppliers, a leakage function $l$ which returns the length of a message and a function $B$ to compute $bl_i^{tp_k}$.       
            \begin{itemize}
            \item Upon receiving an input $("ComputBill",SID,ID_i, m_i^{tp_k},d_i^{tp_k},sk_i^{tp_k},sk_{i,t}^{tp_k})$ from a SM of user $i$, compute $bl_i^{tp_k} = B(m_i^{tp_k},d_i^{tp_k})$, record $(ID_i,bl_i^{tp_k})$ and send a message $(ID_i,l(m_i^{tp_k}+sk_i^{tp_k}), l(sk_{i,t}^{tp_k}))$ to $S$. If user $i$ is corrupted, then disclose $m_i^{tp_k}$ and $d_i^{tp_k}$ to $S$. 

            \end{itemize}
        };
        \draw (textbox.north west) rectangle (textbox.south east);
    \end{tikzpicture}
    \caption{Functionality $\mathcal{F}_{EPIBS}$}
    \label{fig:FEPIBS}
\end{figure}

\begin{definition}[Ideal Functionality of Billing for LEM $\mathcal{F}_{BLEM}$] 
Given a set of input tupples $(ID_i,sID_j,zID_z,v_i^{z,tp_k},d_i^{z,tp_k})$ from users and a set of tupples $(ID_i, zID_z, m_i^{tp_k})$ from their SMs and a set of tupples $(ID_i,b_i^{tp_k})$ from LEMO, the ideal functionality $\mathcal{F}_{BLEM}$ computes 
$BL_i^{bp_u}$ and $SCap_j^{bp_u}$ for all users and suppliers and return the results to the designated parties. 
\end{definition}

Next we show that all three approaches of PBP-LEM securely realise $\mathcal{F}_{BLEM}$. 

\begin{theorem}
    Let $\pi_{\text{App1}}$,$\pi_{\text{App2}}$ and $\pi_{\text{App3}}$ be PBP-LEM employing Approach 1, 2 and 3 , respectively.  Then the protocols $\pi_{\text{App1}}$,$\pi_{\text{App2}}$ and $\pi_{\text{App3}}$ securely emulate the functionality $\mathcal{F}_{BLEM}$ in the \{$\mathcal{F}_{SecComp}$, $\mathcal{F}_{COM}$, $\mathcal{F}_{EPIBS}$\}-hybrid model.
\end{theorem}
\textit{PROOF}. Assuming the presence of concrete implementations of the underlying functionalities $\mathcal{F}_{SecComp}$, such as those in~~\cite{Canetti4,Gilad}, and $\mathcal{F}_{COM}$, such as in~\cite{Canetti3}, the proof is a straightforward application of the composition theorem. Since the protocols $\pi_{\text{App1}}$,$\pi_{\text{App2}}$ and $\pi_{\text{App3}}$ compromise only $\mathcal{F}_{SecComp}$, $\mathcal{F}_{COM}$ and $\mathcal{F}_{EPIBS}$, which can be substituted by their universally composable implementations. 

\paragraph{Authorisation and Collusion Impact Mitigation:} authorisation is achieved by allowing aggregated users' bills to be retrieved by their contracted suppliers only. This is simply accomplished through the use of $sID_j$, which ensures that the relevant parties (KA, LEMO, and CS) forward the individual data necessary for bill computation only to the designated suppliers.

Regarding the collusion impact mitigation, it is achieved by ensuring that any potential collusion between computing parties in the market does not result in the disclosure of critical fine-grained private data (e.g., $b_i^{tp_k}$ and $m_i^{tp_k}$). Specifically, in the case of an honest-majority CS, it is sufficient for two computational servers to collude to reveal private individual data, whereas in the case of a dishonest majority CS, all three servers must collude for malicious collusion to occur. However, in either scenario, the only private data CS are capable of retrieving are $d_i^{tp_k}$ and $v_i^{tp_k}$, which reveal significantly less sensitive information about users compared to $b_i^{tp_k}$ and $m_i^{tp_k}$. Additionally, collusion between CS and suppliers has no higher impact since all private data in the possession of users' suppliers is encrypted (using EPIBS and FHIPE).

\section{Performance Evaluation}\label{sec:evaluation}
This section evaluates the performance of PBP-LEM in terms of computational and communication complexity, while considering the three proposed approaches. 

\subsection{Computational Complexity}
The most computationally demanding operations in PBP-LEM are IPE methods (specifically IPE.LeftEncrypt, IPE.RightEncrypt, and IPE.Decrypt), multiplication, and comparison operations performed by CS, followed by Pedersen commitment methods (Commit and Open). The computation cost of all remaining operations, including EPIBS operations, is negligible.
\begin{table}[t]
  \centering
  \scriptsize
  \caption{Computational Complexity.}
\label{Computation-Results}
\begin{tabular}{ l @{\hspace{0.9\tabcolsep}}  | c @{\hspace{0.9\tabcolsep}} | c @{\hspace{0.9\tabcolsep}} | c @{\hspace{0.9\tabcolsep}} |c @{\hspace{0.9\tabcolsep}} } 
\toprule
  \multirow{2}{*}{\textbf{Entity}}  &\multicolumn{3}{c}{\textbf{\shortstack{Per $tp_k$}}}  &  \shortstack{\textbf{Per $bp_u$}}\\
       
        \noalign{\vspace{0.5ex}}\cline{2-5} \noalign{\vspace{0.5ex}}\\
        & All Apprs. & Only Appr. 1 & Only Appr. 2 & All Apprs. \\
      
        \midrule
          SM & $Commit$ & $N_v * IPE.LeftEncrypt$ & - & -\\
          User & $Commit$ & $2 * N_v * IPE.RightEncrypt$ & - & -\\
          Supplier & $N_i^j * 2 * HomoAdd$ & $N_i^j * 2 * N_v * IPE.Decrypt $& - &  $ N_i^j * Open $\\
           CS & - & - & $N_i \times Comp + N_i \times Mult$ & - \\
         
\bottomrule

\end{tabular}
\end{table}

\begin{figure}[t]
\centering
\includegraphics[width=0.95\columnwidth,trim=4 4 4 4,clip]{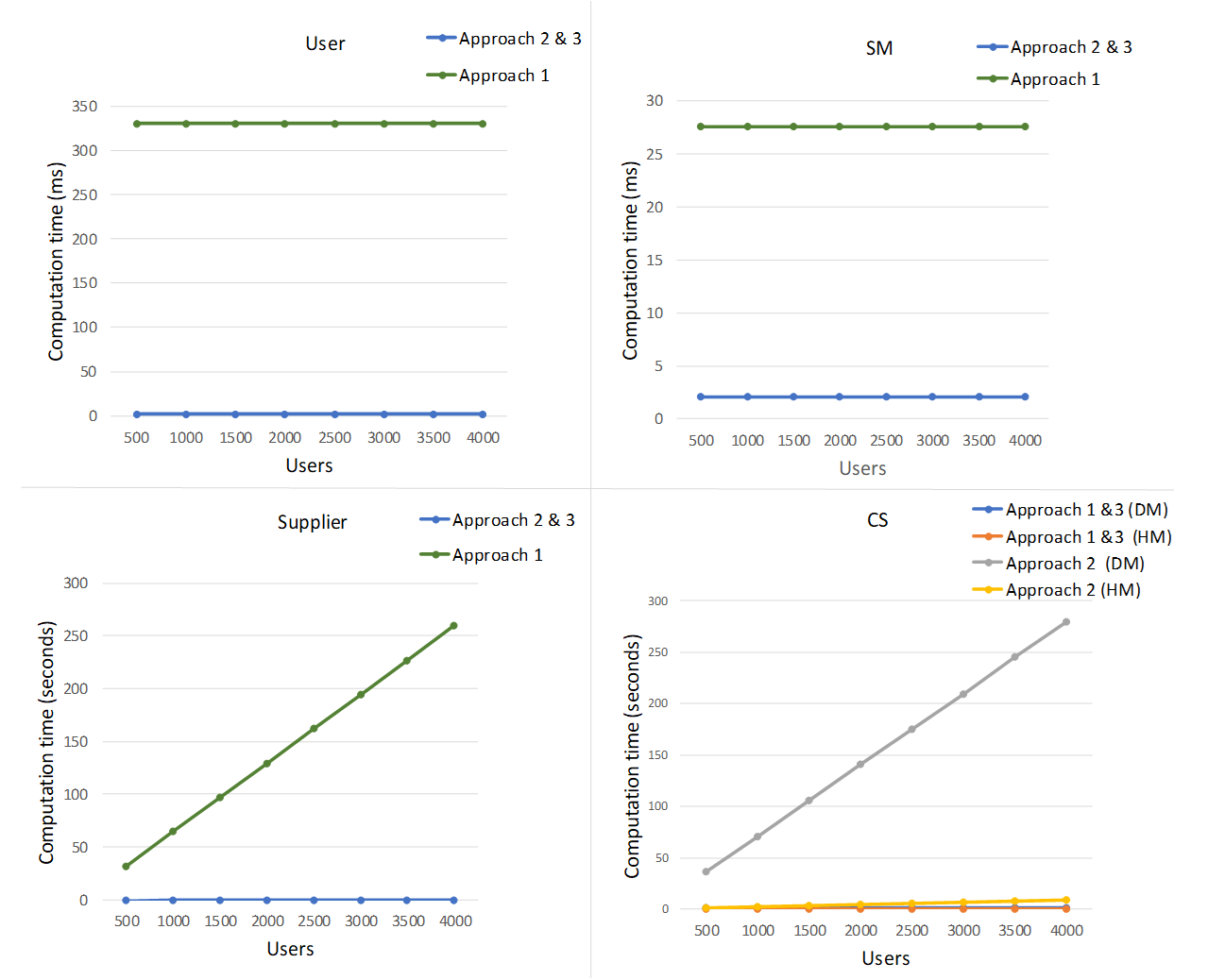}
  \captionsetup{justification=centering}
\caption{Computational overhead for each entity per trading period.}
\label{ComputationalOverhead}
\end{figure}

For all approaches, for every $tp_k$, each SM computes one Commit over its recorded meter reading; each user computes one Commit over its bid volume and each supplier computes two homomorphic additions on the commitments for each of its contracted user: $N_u^s \times 2 \times $ HomoAdd. Moreover, for every $bp_u$, every supplier performs one commitment opening for each corresponding user:  $N_u^s \times$ Open. If approach (1) is adopted, in addition to the previously mentioned operations, each SM computes $N_v \times$ IPE.LeftEncrypt; every user computes $N_v \times 2 \times $ IPE.RightEncrypt and each supplier computes $N_u^s \times N_v \times 2 \times $ IPE.Decrypt in the ultimate worst case scenario. Here, $N_v$ is the number of vectors required to encode each bid volume and meter reading using dual binary encoding scheme methods (i.e. $f_y$ and $f_x$)~\cite{Gaybullaev}. Please note that suppliers may perform $N_v \times 2 \times $ IPE.Decrypt for each user in the worst-case scenario, which occurs when the bid volume and meter reading values are equal. However, in most cases, the values result in a better cost, as the ``for" loop in Algorithm(\ref{alg:IPEComparision}) may exit before iterating $N_v$ times. In the case of approach (2), CS additionally performs one comparison and one multiplication for each user in the market: $N_u \times Comp + N_u \times Mult$. In case approach (3) is employed, no additional computationally heavy operations are to be performed. The computational complexity of PBP-LEM is summarised in Table~\ref{Computation-Results}.

We also conducted experiments to test the performance of PBP-LEM. CS, supplier and KA were implemented on a 64-bit Linux server equipped with 16 cores single thread Intel Xeon processors and 64 GB of memory. We run all the parties on the same server, therefore, network latency is not considered. Users and SMs operations were evaluated on a laptop with Intel Core i7 CPU and 8 GB of memory. CS algorithms were executed using the MP-SPDZ framework~\cite{Keller2}, which supports the underlying primitives and MPC protocols employed by PBP-LEM (Section~\ref{Sec:MPC}). All other entities' algorithms were implemented in python. IPE modules was adopted from ~\cite{FHIPE}.

We followed the same random generation approach used in~\cite{Madhusudan} --based on a realistic dataset from~\cite{Abidin} -- to simulate bid volumes and meter readings over a 30-minute trading period and one month billing period. RP, FiT and TP are set to .3£/Wh, .1£/Wh and .2£/Wh, respectively.  We initiated our experiments with 1000 users participating in a LEM for one trading period and gradually increased the number to 4000 users. The user distribution comprised an equal split between prosumers and consumers. We assumed that there are 6 different suppliers and four zones in the market, allocating users equally among them. 

Based on the generated bid volumes and meter reading values ranges, we set $N_v$ to 12 for Approach (1), allowing the representation of 4095 different integer values according to the dual binary encoding scheme~\cite{Gaybullaev}. Our results indicate that each Commit costs 2.10 ms; HomoAdd costs 0.124812 ms; Open costs 0.25 ms; $N_v \times $ IPE.LeftEncrypt costs 25.40 ms; $N_v \times $ IPE.RightEncrypt costs 164.10 ms; {.$2 \times N_v \times $ IPE.Decrypt costs 622 ms}.

Figure~\ref{ComputationalOverhead} illustrates the computational cost per $tp_k$ for each entity in the market with respect to each of the three approaches. Approach (1) incurs additional costs for the resource-constrained market entities (i.e., SMs and users) compared to the other approaches, as it involves more intensive operations during Step (1). Nevertheless, none of the three approaches impose significant computational demands on SMs and users. For instance, the overall computational time per $tp_k$ for each SM and user is 2.10 ms in Approaches (2) and (3), and 27.56 ms for SM and 330.14 ms for user in Approach (1). Regarding suppliers and CS, the selected approach during Step (2) determines which entity should handle the most computationally intensive operations to incorporate deviation costs in bill calculations. This costs arises from Algorithm (\ref{alg:SPCSComparision}) performed by CS in approach (2) or Algorithm (\ref{alg:IPEComparision}) executed distributively by users' respective suppliers in approach (1). Please note that, as we evenly distributed the number of users among suppliers, they all share similar costs. However, in a more realistic scenario, some suppliers might bear higher costs than others based on the number of their contracted users. In approach (3), users' privacy has been relaxed in exchange for eliminating heavy computations, resulting in a trivial cost for both suppliers and CS.
\begin{figure}[t]
\centering
\includegraphics[width=0.5\columnwidth,trim=4 4 4 4,clip]{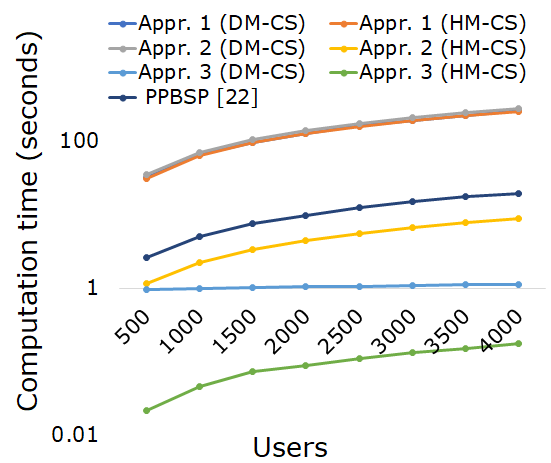}
  \captionsetup{justification=centering}
\caption{Overall bills computation cost per trading period.}
\label{ComputationalOverhead2}
\end{figure}

Figure~\ref{ComputationalOverhead2} displays the total computation time for bill calculation per $tp_k$ in all approaches, as well as PPBSP by~\cite{Andrei} for benchmarking. Please note that the figure accounts for CS's computation time, considering the offline phase of the underlying MPC primitives. Clearly, Approach (3) stands out as the most efficient, attributed to privacy loosening—specifically, disclosing individual deviations to the suppliers. If such leakage is not tolerated by users, the decision between approaches (1) and (2) becomes less straightforward, given our parameter assumptions. In the case of a malicious majority CS, the overall computational time for bill computation in approaches (1) and (2) is comparable, and the key difference lies in the part of the market where the cost is concentrated, i.e., CS or suppliers. However, considering only the online phase of CS would give preference to approach (2). Another crucial factor to consider is the number of suppliers in the market and the distribution of users among them. For instance, assuming that over 20\% of market users are associated with a single supplier would result in a slower bill computation time for those users in Approach (1) compared to Approach (2). In the scenario of an honest-majority CS, Approach (2) is preferable, providing a more private solution than Approach (3) and significantly greater efficiency than Approach (1).

Compared to PPBSP, PBP-LEM excels only in the case of Approach (3) and in Approach (2) with an honest-majority CS, where the former requires around 0.18 seconds and the later 8.90 seconds to compute the bills for 4000 users per $tp_k$, including the offline phase. Nevertheless, bill computation for the other approaches can be completed maximally in less than 5 minutes for 4000 users per $tp_k$, demonstrating the feasibility of PBP-LEM for real LEM deployment.

\begin{table}[t]
  \centering
  \scriptsize
  \begin{threeparttable}
  \caption{Communication Complexity.}
\label{Communication-Results}
\begin{tabular}{ l @{\hspace{0.9\tabcolsep}}  | c @{\hspace{0.9\tabcolsep}} | c @{\hspace{0.9\tabcolsep}} | c @{\hspace{0.9\tabcolsep}} |c @{\hspace{0.9\tabcolsep}} } 
\toprule
  \multirow{2}{*}{\textbf{Entity}}  &\multicolumn{3}{c}{\textbf{\shortstack{Per $tp_k$}}}  &  \shortstack{\textbf{Per $bp_u$}}\\
       
        \noalign{\vspace{0.5ex}}\cline{2-5} \noalign{\vspace{0.5ex}}\\
        & All Apprs. & Only Appr. 1 & Only Appr. 2 & All Apprs. \\
      
        \midrule
          SMs-to-Suppliers &  $N_i * (|C|$ + $|\langle X \rangle)|$ & $N_i * N_v * |CT_l|$ & - & $N_i * |R|$\\ 
         LEMO-to-Suppliers & $N_i * (|\langle X \rangle|+ |C|)$ & $2 * N_i * N_v * |CT_r|$ & - & -\\
         Users-to-CS & $ 6 * N_i * |[X]|$& & - & - \\
         CS-to-Suppliers & - & - & $3 *N_i * |[X]| $&$N_i * |X|$\\ 
         Suppliers-to-Users  & - & - & - &$N_i * |X|$\\ 
         Suppliers-to-KA & $2 * N_i * |CN|$  & - & - & -\\
         KA-to-SMs/Users  & - & - & - & $N_i * (2880 * |K| + |MK|)$  \\ 
          KA-to-Suppliers  & - & - & - & $N_i * (4 * |K * X| + |pp|)$\\ 
\bottomrule

\end{tabular}
\end{threeparttable}
\end{table}
\begin{figure}[t]
\centering
\includegraphics[width=0.95\columnwidth,trim=4 4 4 4,clip]{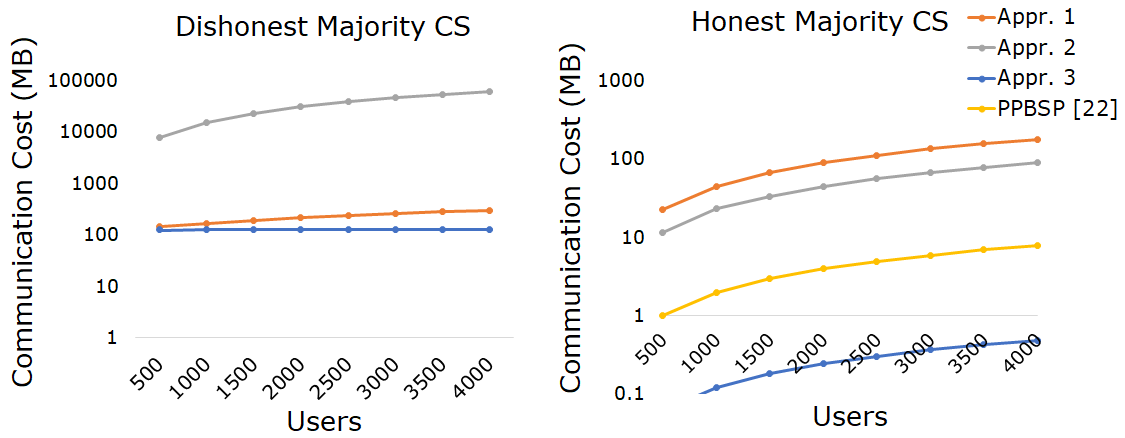}
  \captionsetup{justification=centering}
\caption{Total communication cost per trading period.}
\label{CommunicationOverhead}
\end{figure}
\subsection{Communication Complexity}
In evaluating communication complexity, we focus specifically on the costs associated with implementing privacy techniques and measures within PBP-LEM. Table~\ref{Communication-Results} provides a summary of the communication complexity in PBP-LEM for different approaches, where, $|X|$ represents the size of a single value, $|[X]|$ the size of a share, $|<X>|$ the size of a Pedersen commitment, $|C|$ the size of a ciphertext using EPIBS, $|R|$ the size of a random value used in Pedersen commitments, $|CT_l|$ the size of a ciphertext using $IPE.LeftEncrypt$, $|CT_r|$ the size of a ciphertext using $IPE.RightEncrypt$, $|CN|$ a boolean value, $|K|$ the size of a random key generated for EPIBS, and $|MK|$ is the size of the master key generated for FHIPE. $|CT_l|$ and $|CT_r|$ have been evaluated to be 550 and 1650 bytes, respectively, considering that the number of elements in each vector is $N_v + 1$, i.e., 13. We measure the communication cost between CS using MP-SPDZ and set $|[X]| = 64$, $|<X>| = |R|= 256$, $ |K| = 128$, $|X| = |C| = 32$, $ |CN| = 1$ bits.

Figure~\ref{CommunicationOverhead} presents the communication cost of PBP-LEM and PPBSP~\cite{Andrei}. PBP-LEM outperforms PPBSP only in the case of the relaxed privacy approach, i.e., Approach 3. Comparing PBP-LEM's approaches, as expected, Approach 3 exhibits the significantly lowest communication cost due to the disclosure of individual deviations. This indicates that the major communication overhead arises from the protection of individual deviations that requires private comparisons. With regards to the choice between Approach 2 and Approach 3, the decision depends on the chosen security model of the CS. For an honest-majority CS, the communication cost of Approach 2 is almost half that of Approach 1. However, in the case of a dishonest-majority CS, Approach 2 incurs a significantly higher cost concentrated between CS parties.

\section{Conclusion}\label{sec:conclusion}
We introduced a privacy-preserving billing protocol for LEMs (PBP-LEM). PBP-LEM considers imperfect bid fulfillment and distributes bill computation among different parties (i.e., CS and suppliers) while protecting users' private data (e.g., bid volumes, smart meter readings, deviations, and locations) and mitigating collusion impact. Furthermore, we presented three approaches that offer varying degrees of privacy and performance. We also proved that PBP-LEM fulfill its security and privacy requirements. Finally, we analysed the associated computation and communication costs of all our approaches under different security settings. The results demonstrate the variations in performance between the approaches and indicate the feasibility of PBP-LEM for a setting based on real LEMs.

\renewcommand{\bibname}{References}
\bibliographystyle{splncs04}
\bibliography{Refs}
\newpage
\appendix
\begin{landscape}
\section{PBP-LEM Overview}\label{App:Overview}
\begin{figure}
\includegraphics[width=.7\columnwidth]{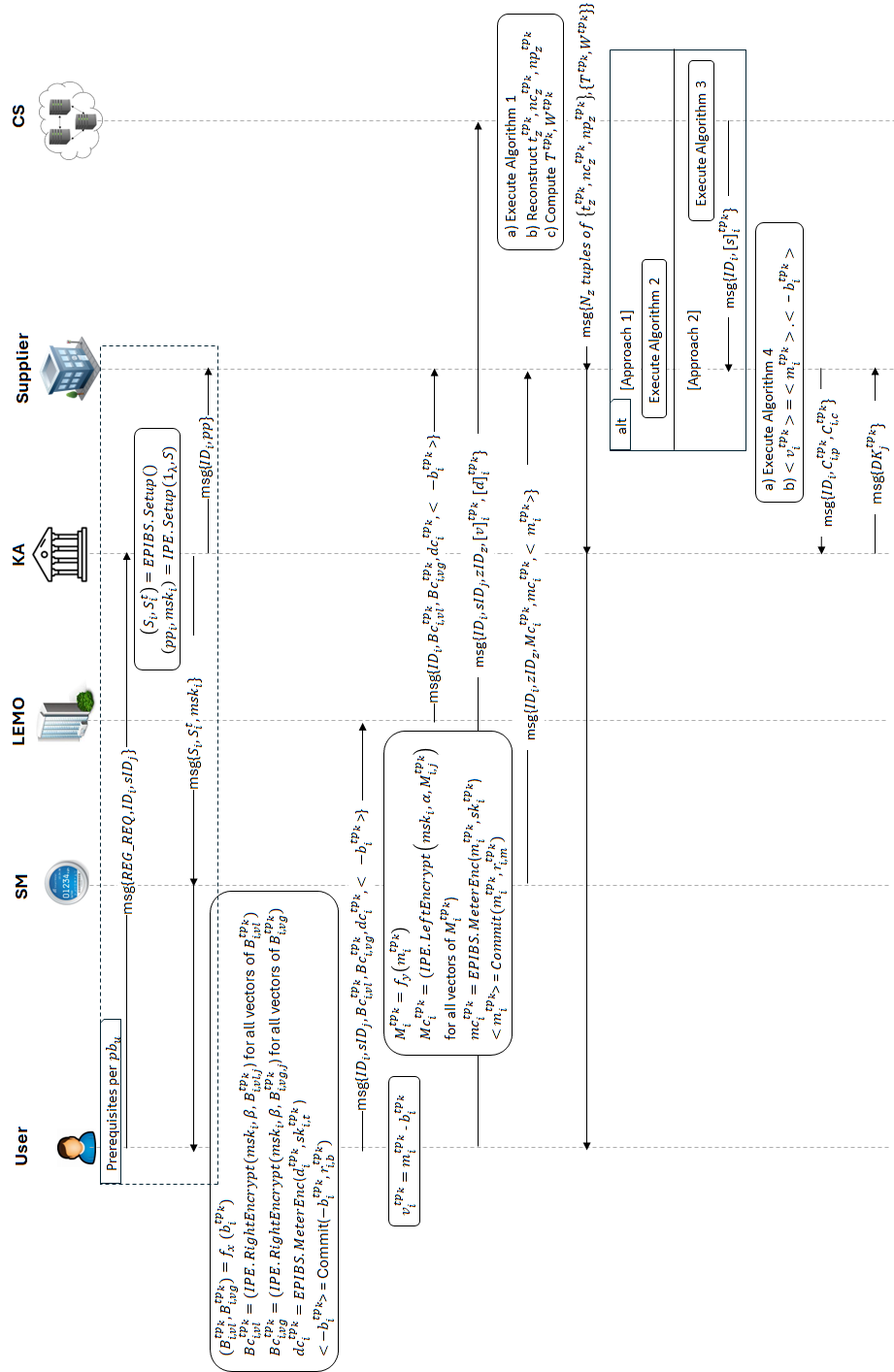}
  \captionsetup{justification=centering}
\caption{PBP-LEM Prerequisites and Overview Per Trading Period.}
\label{OverviewPT}
\end{figure}
\begin{figure}
\includegraphics[width=.66\columnwidth]{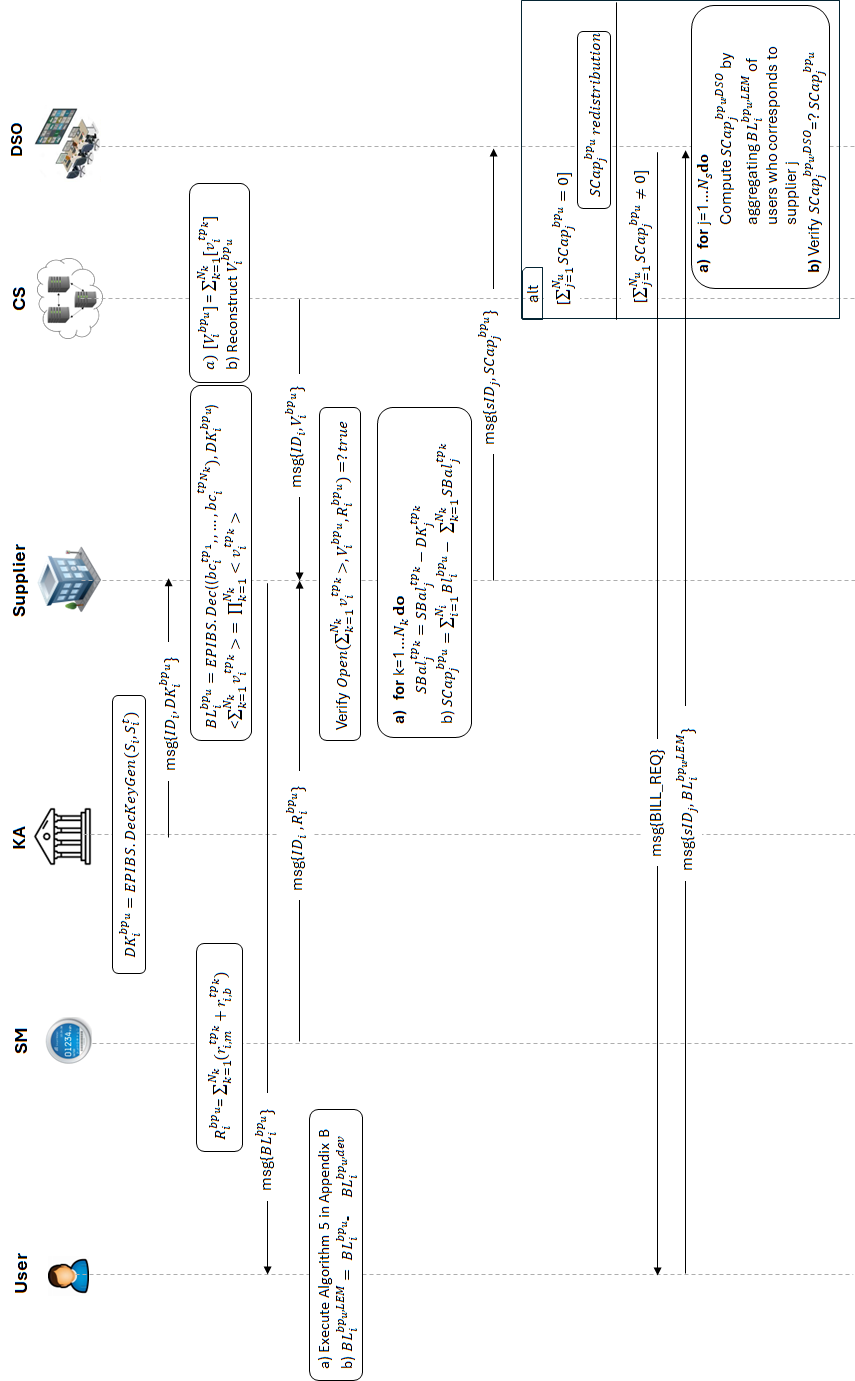}
  \captionsetup{justification=centering}
\caption{PBP-LEM Overview Per Billing Period.}
\label{OverviewPB}
\end{figure}
\section{Bill Computation in Clear by Users Per Billing Period}\label{App:BillsComputation}

 \begin{algorithm}[t]
      \footnotesize
        \caption{Bill Computation by Users Per Billing Period}\label{alg:BillsComputation}
        \hspace*{\algorithmicindent} \textbf{Input:}$m_i^{tp_k}$, $d_i^{tp_k}$, $v_i^{tp_k}$, ($T^{tp_k}, W^{tp_k}$), zone z tuple $ZN =  (t_z^{tp_k}, nc_z^{tp_k}, np_z^{tp_k})$ to which the user belongs.\\
        \hspace*{\algorithmicindent} \textbf{Output:}  $Bl_i^{bp_u,L},Bl_i^{bp_u,LEM}$
        \begin{algorithmic}
        \For{$k = 0$ to $N_k$ }
            \State $Bl_i^{bp_u,LEM} += m_i^{tp_k} * TP^{tp_k}$
            \State $Bl_i^{bp_u,L} += Bl_i^{bp_u,LEM}$
           \If{$T^{tp_k} >0$ and $t_z^{tp_k} >0$ and $v_i^{tp_k} >0$ and $d_i^{tp_k} =1$}
              \State $Bl_i^{bp_u,LEM} -= TP^{tp_k} * (t_z^{tp_k} * W^{tp_k}/ np_z^{tp_k})$
              \State $Bl_i^{bp_u,L} = Bl_i^{bp_u,LEM} + FiT^{tp_k} * (t_z^{tp_k} * W^{tp_k}/ np_z^{tp_k})$
            \ElsIf{$T^{tp_k} <0$ and $t_z^{tp_k} <0$ and $v_i^{tp_k}< 0$ and $d_i^{tp_k} = 0$}
               \State $Bl_i^{bp_u,LEM} -= TP^{tp_k} * (t_z^{tp_k} * W^{tp_k}/ np_z^{tp_k})$
               \State $Bl_i^{bp_u,L} = Bl_i^{bp_u,LEM} + RP^{tp_k} * (t_z^{tp_k} * W^{tp_k}/ np_z^{tp_k})$
               
            \EndIf
        \EndFor
        \end{algorithmic}
\end{algorithm}

\end{landscape}
\end{document}